\documentclass[aps,twocolumn,showpacs,superscriptaddress,groupedaddress,nofootinbib]{revtex4-1} 
\usepackage{graphicx}
\usepackage[sort&compress]{natbib}
\usepackage{lipsum}
\usepackage{morefloats}
\usepackage[pdf]{pstricks}
\usepackage{subfigure}
\usepackage{amsmath}
\usepackage{amssymb}
\usepackage{amsfonts}
\usepackage{rotating}
\usepackage{cancel}
\usepackage{mathtools}
\usepackage{color}
\usepackage{bbm}
\usepackage{dsfont}
\usepackage{bbold}
\usepackage{multirow}
\usepackage{ulem}
\newcommand{\eps}{\epsilon}

\begin{document}

\title{$\Lambda(1405)$ mediated triangle singularity in the $K^-d\to p\Sigma^-$ reaction}
\author{A. Feijoo}
\email{edfeijoo@ific.uv.es}
\affiliation{Departamento de F\'{\i}sica Te\'orica and IFIC,
Centro Mixto Universidad de Valencia-CSIC,
Institutos de Investigaci\'on de Paterna, Aptdo. 22085, 46071 Valencia, Spain}
\affiliation{Nuclear Physics Institute, 25068 Rez, Czech Republic}   
\author{R. Molina}
\email{Raquel.Molina@ific.uv.es}
\affiliation{Departamento de F\'{\i}sica Te\'orica and IFIC,
Centro Mixto Universidad de Valencia-CSIC,
Institutos de Investigaci\'on de Paterna, Aptdo. 22085, 46071 Valencia, Spain}  
\author{L. R. Dai}
\email{dailianrong@zjhu.edu.cn}
\affiliation{School of Science, Huzhou University, Huzhou 313000, Zhejiang, China}
\author{Eulogio Oset}
\email{Eulogio.Oset@ific.uv.es}
\affiliation{Departamento de F\'{\i}sica Te\'orica and IFIC,
Centro Mixto Universidad de Valencia-CSIC,
Institutos de Investigaci\'on de Paterna, Aptdo. 22085, 46071 Valencia, Spain}

\begin{abstract}  
We study for the first time the $p\Sigma^-\to K^-d$ and $K^-d\to p\Sigma^-$ reactions close to threshold and show that they are driven by a triangle mechanism, with the $\Lambda(1405)$, a proton and a neutron as intermediate states, which develops a triangle singularity close to the $\bar{K}d$ threshold. We find that a mechanism involving virtual pion exchange and the $K^-p\to\pi^+\Sigma^-$ amplitude dominates over another one involving kaon exchange and the $K^-p\to K^-p$ amplitude. Moreover, of the two $\Lambda(1405)$ states, the one with higher mass around $1420$ MeV, gives the largest contribution to the process. We show that the cross section, well within measurable range, is very sensitive to different models that, while reproducing $\bar{K}N$ observables above threshold, provide different extrapolations of the $\bar{K}N$ amplitudes below threshold. The observables of this reaction will provide new constraints on the theoretical models, leading to more reliable extrapolations of the $\bar{K}N$ amplitudes below threshold and to more accurate predictions of the $\Lambda(1405)$ state of lower mass.
\end{abstract}
\maketitle
\section{Introduction}
Introduced early in the 50's \cite{1,2}, the triangle singularities (TS) are getting a growing attention nowadays since they are helping to understand many phenomena observed in hadron physics. The singularity stems from a mechanism that can be depicted by a triangle Feynman diagram, see Fig. 1, where a particle A decays into 1 and 2, 1 decays into B and 3, and 2 and 3 merge to form particle C. If this mechanism can occur at the classical level, a singularity appears in the amplitude (Coleman-Norton theorem \cite{3}), which requires that 1 and B move in the same direction in the A rest frame, and 3 moves in the direction of 2 and faster such that it catches up with 2 and fuses with it to give C. The subject has been reformulated recently with a more intuitive and practical formalism in Ref. \cite{4} and a thorough review has been done in Ref. \cite{23}.

\begin{figure}
 \centering
 \includegraphics[scale=0.7]{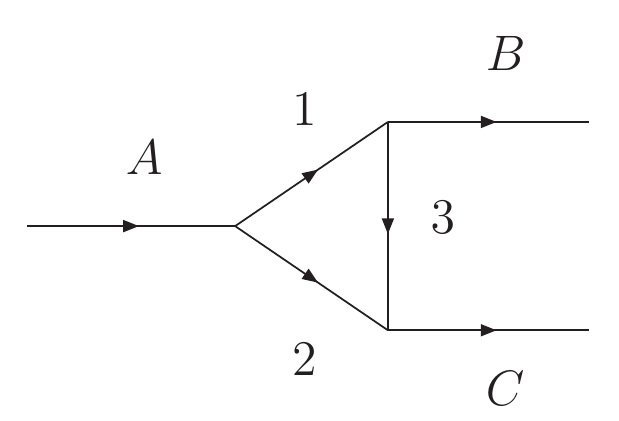}
 \caption{Feynman diagram from where a triangle singularity can emerge. Particle A decays into 1 and 2, 1 decays into B and 3, and 2 and 3 merge to form particle C.}
 \label{fig:1}
\end{figure}

Recent examples of TS are found in the study of the $\eta(1405)\to f_0(980) \pi^0$ decay \cite{7} performed in Refs. \cite{8,9,10,11}. Another relevant case was the explanation of the ``$a_1(1420)$'' structure observed by the COMPASS collaboration \cite{15}, which is explained in terms of a TS in Refs. \cite{16,17,18,19}. Some other recent examples can be seen in \cite{20,21,22} and a rather complete list of reactions studied along TS is given in Ref. \cite{23}.

Another example of TS is given by the $\pi^+d\to pp$ reaction \cite{32,said} which has been much studied in the past \cite{35,36,37,39}. Recently, this latter reaction got again attention in \cite{ikenoraquel, raquelikeno} by looking at the time reversal reaction $pp\to \pi^+d$, because it was shown to be driven by a triangle mechanism with $\Delta NN'$ in the intermediate states and $NN'$ fusing to give the deuteron. This mechanism was found responsible for the relatively large cross section of the fusion reaction. The relevance of this reaction was stressed by the fact that the sequential one pion production process, $pn(I=0)\to\pi^-pp\to \pi^-\pi^+d$ (plus $pn(I=0)\to \pi^+nn\to\pi^+\pi^-d$) produced a peak in the cross section \cite{raquelikeno} with the right strength and width around $\sqrt{s}=2350$ MeV which provided a natural explanation of the experimental peak seen in the $pn\to \pi^+\pi^-d$ and $pn\to \pi^0\pi^0d$ reactions around this energy  \cite{25,26}, which so far has been attributed to a dibaryon $d^*(2380)$. In addition to the properties of the peak generated by this reaction, the nature of the TS gave rise to a particular structure that favored the quantum numbers of the peak observed as $I(J^P)=0(3^+)$, as has been found experimentally \cite{31}.

In the present case we shall study the related $p\Sigma^-\to K^-d$ ($K^-d\to p\Sigma^-$) reactions, showing that the mechanism is similar to the one of the $pp\to \pi^+d$ reaction, but now the high energy pole of the $\Lambda(1405)$ plays the role of the $\Delta(1232)$ in the $pp\to \pi^+d$ reaction. The TS places this $\Lambda(1405)$ and the nucleons of the triangle diagram on shell and this leads to the interesting result that one can see the effects of the $\Lambda(1405)$, which lies below the $\bar{K}N$ threshold, in a reaction with physical kaons, in other words, we observe effects of the $K^-p$ amplitudes below threshold in a reaction with $K^-N$  above threshold. This is most welcome, since different theoretical models for the $\bar{K}N$ interaction reproducing well the data above the $\bar{K}N$ threshold lead to quite different results for the amplitudes below threshold. 

The $\bar{K}N$ interaction has been the subject of intense theoretical scrutiny \cite{dalitz, thomas, fink, rubin} which has been reinforced with the advent of the chiral unitary approach \cite{38M,39M,40M, 44M, Lutz, 46M}. One surprise from the use of this approach is the existence of two $\Lambda(1405)$ states \cite{44M,47M} which have found its way into the PDG \cite{pdg} only recently. A large amount of papers have come to corroborate this finding \cite{48M, 49M, 50M, 54M, 56M, luisone, 58M, 59M, 61M, hyodoweise, 67M, 68M, miyahara, hyodojido}. Reviews on this issue can be seen in \cite{hyodoulf, 29M, symmetry}. Yet, in spite of reproducing the $\bar{K}N$ data above threshold and some threshold observables, the different models produce $\bar{K}N$ amplitudes below threshold which differ much from each other (see Fig. 1 of \cite{cieplyconf}). We will show that because the reaction relies upon $\bar{K}N$ amplitudes below threshold, the results that we obtain with several models for the $\bar{K}N$ interaction lead to results for the $K^-d\to p\Sigma^-$ reaction that differ appreciably among themselves. In other words, the measurement of this cross section would provide an extra valuable observable to put more constraints on the theoretical models and make them more predictive below the $\bar{K}N$ threshold. This information would go in the same line of the work presently done at DAFNE in the programs as AMADEUS \cite{DelGrande:2021wpg} and SIDDHARTA \cite{Curceanu:2020kkg}.

\section{Formalism}
\begin{figure*}
\centering
 \includegraphics[width=0.85\textwidth]{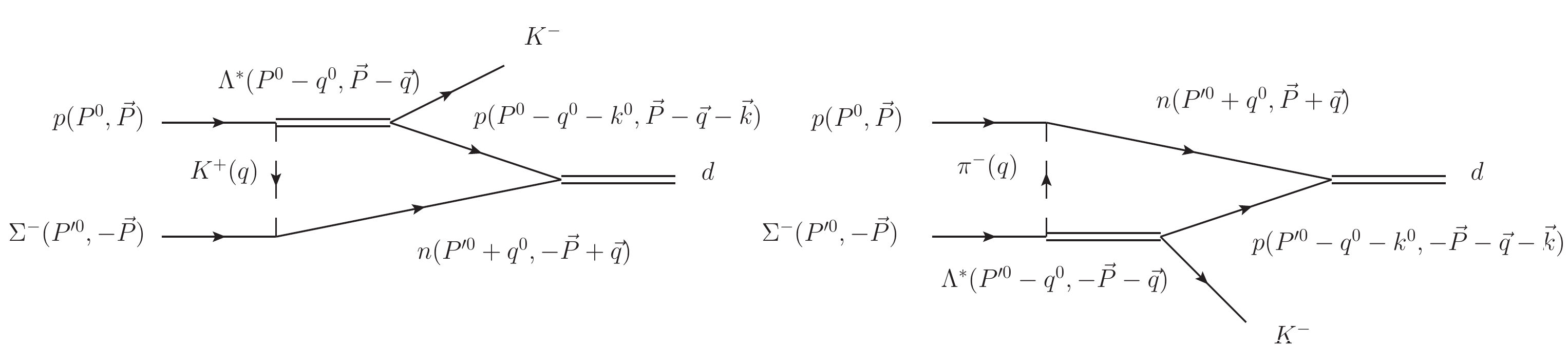}
 \caption{Feynman diagrams for the $p\Sigma^-\to K^-d$ reaction. The momenta of the lines are shown in brackets.
}
 \label{fig:2}
 \end{figure*}

We shall study the $p\Sigma^-\to K^-d$ reaction to be able to exploit the analogies with the $pp\to \pi^+d$ reaction. The process proceeds via the diagrams of Fig. \ref{fig:2}.

The diagrams of Fig. \ref{fig:2} develop a triangle singularity. This occurs when the $\Lambda(1405)$, the $p$ and $n$ intermediate states are placed on shell in the loop (the $K^+$ and $\pi^-$ are off shell and do not matter for the discussion of the TS) and the $\Lambda(1405)$ and $K^-$ are in the same direction. If the proton, which goes in the same direction of the neutron in this case, goes faster than the neutron it can catch up with the neutron and fuse into the deuteron, producing the TS according to the Coleman-Norton theorem \cite{3}. All these conditions are summarized in Eq. (18) of Ref. \cite{4} in the limit of zero width of the $\Lambda(1405)$,
\begin{equation}
 q_\mathrm{on}=q_{a^-}\label{7.1}
\end{equation}
where $q_{\mathrm{on}}$ is the momentum of the neutron in the $p\Sigma^-$ rest frame and $q_{a^-}$ is one of the solutions of the momentum of the neutron in the decay of the $d$ into $pn$ for the moving $d$ in the $p\Sigma^-$ rest frame. Easy analytical formulae for $q_{\mathrm{on}}$, $q_{a^-}$ are given in Ref. \cite{4}. Technically, the deuteron bound does not decay to $pn$, and to test Eq. (1) one can take a deuteron slightly unbound. The singularity becomes a broad peak upon the consideration of the $\Lambda(1405)$ width and, by continuity,  it shows up even if the deuteron is bound by $2.2$ MeV. The test of Eq. (1) tells that the TS should be around $\sqrt{s}=2380$ MeV, a bit above the $K^-d$ threshold. Technically one would perform the $K^-d\to p\Sigma^-$ measurement, which can be done for low $K^-$ energies in DAFNE \cite{catalina} and in other facilities, as JPARC \cite{Sato:2009zze,Kumano:2015gna} or the planned kaon facility at Jefferson Lab \cite{Briscoe:2015qia,Amaryan:2020xhw}.

To evaluate the amplitudes for the diagrams of Fig. \ref{fig:2} one needs the coupling of the two $\Lambda(1405)$ to $K^-p$ and to $\pi^+\Sigma^-$ and the couplings of $\pi^-p\to n$ and $K^+\Sigma^- \to n$. The $\pi^-pn$ coupling is given for an incoming $\pi^-$ of momentum $\vec{q}$ by 
\begin{equation}
 -it=\frac{f_{\pi NN}}{m_\pi}\vec{\sigma}\cdot \vec{q}\sqrt{2}\ ,\label{eq:fpinn}
\end{equation}
with $f_{\pi NN}=1.002$. Alternatively, this coupling and in general the pseudoscalar baryon vertex (PBB), can be obtained from chiral labrangians \cite{ecker,ulfito} and the general result is given by \cite{angelsphi},
\begin{eqnarray}
 -it_{\bar{K}NY}=\left[\alpha\frac{D+F}{2f}+\beta\frac{D-F}{2f}\right]\vec{\sigma}\cdot\vec{q}\label{eq:tkny}
\end{eqnarray}
for an incoming $\bar{K}$, with $f=93$ MeV, $D=0.795$, $F=0.465$ \cite{borasoy}. In particular,
\begin{equation}
 -it_{K^+\Sigma^-\to n}=\sqrt{2}\frac{D-F}{2f}\vec{\sigma}\cdot\vec{q}\label{eq:tksn}
\end{equation}
We note that in this formalism, $\frac{f_{\pi NN}}{m_\pi}=\frac{D+F}{2f}$. The isospin $I=0$ function for the deuteron can be written as
\begin{equation}
 \vert d\rangle=\frac{1}{\sqrt{2}}(\vert pn\rangle-\vert np\rangle) .
\end{equation}
The coupling of the deuteron to $NN$ is given by $g_d$. In particular for the $pn$ component 
\begin{eqnarray}
 -it_{dpn}=-ig_d\frac{1}{\sqrt{2}}\theta(q_\mathrm{max}-\vert \vec{q}\vert_{\mathrm{c.m.}})
\end{eqnarray}
where $\vec{q}_{\mathrm{c.m}}$ is the $p$ momentum in the $d$ rest frame. From the study in Appendix A of \cite{ikenoraquel}, we find
\begin{eqnarray}
 &&g_d=(2\pi)^{(3/2)}2.67\times 10^{-3} \mathrm{MeV}^{-1/2} ,\nonumber\\
 &&q_\mathrm{max}=240\,\mathrm{MeV} .
\end{eqnarray}
With these ingredients we can evaluate the amplitudes corresponding to the two diagrams of Fig. \ref{fig:2} and find
\begin{widetext}
\begin{align}
 &-it^{(a)}= (-i)g_{\Lambda^*,K^-p}(-i)g_{\Lambda^*,K^-p}(-i) g_d\frac{D-F}{2f}\int \frac{d^4q}{(2\pi)^4}\vec{\sigma}_2\vec{q}\frac{i}{q^2-m^2_K+i\eps}\frac{M_{\Lambda^*}}{E_\Lambda^*}\frac{i}{P^0-q^0-E_\Lambda^*(\vec{P}-\vec{q})+i\frac{\Gamma_{\Lambda^*}}{2}}\nonumber\\&\times \frac{M_{N}}{E_{N}}\frac{i}{P^0-q^0-k^0-E_{N}(\vec{P}-\vec{q}-\vec{k})+i\eps}\frac{M_N}{E'_N}\frac{i}{P'^0+q^0-E'_N(-\vec{P}+\vec{q})+i\eps}\theta(q_\mathrm{max}-\vert \vec{P}-\vec{q}-\frac{\vec{k}}{2}\vert),\nonumber\\
 &-it^{(b)}=-g_{\Lambda^*,K^-p}g_{\Lambda^*,\pi^+\Sigma^-}g_d\frac{f_{\pi NN}}{m_\pi}i\int\frac{d^4q}{(2\pi)^4}\frac{1}{q^2-m^2_\pi+i\eps} \frac{M_{\Lambda^*}}{E_{\Lambda^*}}\frac{1}{P'^0-q^0-E_{\Lambda^*}(-\vec{P}-\vec{q})+i\frac{\Gamma_{\Lambda^*}}{2}}\nonumber\\&\times\vec{\sigma}_1\cdot \vec{q}\,\frac{M_N}{E_N}\frac{1}{P'^0-q^0-k^0-E_N(-\vec{P}-\vec{q}-\vec{k})+i\eps}\frac{M_N}{E'_N}\frac{1}{P^0+q^0-E'_N(\vec{P}+\vec{q})+i\eps}\theta(q_{\mathrm{max}}-|-\vec{P}-\vec{q}-\frac{\vec{k}}{2}|),
 \end{align}
\end{widetext}
where $E(\vec{P})=\sqrt{\vec{P}^2+m^2}$. 
We write the pseudoscalar propagator as
\begin{eqnarray}
 \frac{1}{\vec{q}\,^2-m^2+i\epsilon}=\frac{1}{2\omega(\vec{q})}\left(\frac{1}{q^0-\omega(\vec{q})+i\epsilon}-\frac{1}{q^0+\omega(\vec{q})-i\epsilon}\right)\nonumber\\\nonumber\\
\end{eqnarray}
with $\omega(\vec{q})=\sqrt{m^2+\vec{q}\,^2}$, and then perform the $q^0$ integration analytically. However, it is practical to reduce the number of denominators containing $q^0$ and for this purpose it is useful to write
\begin{eqnarray}
 &&\frac{1}{P^0-q^0-k^0-E_N(\vec{P}-\vec{q}-\vec{k})+i\epsilon}\nonumber\\&&\times\frac{1}{P^{'0}+q^0-E_N'(-\vec{P}+\vec{q})+i\epsilon}\nonumber\\&&=\frac{1}{\sqrt{s}-k^0-E_N(-\vec{P}+\vec{q})-E_N(\vec{P}-\vec{q}-\vec{k})+i\epsilon}\nonumber\\
 \nonumber\\&&\times\left(\frac{1}{P^0+q^0-k^0-E_N(\vec{P}-\vec{q}-\vec{k})+i\epsilon}\right.\nonumber\\&&\left.+\frac{1}{P^{'0}+q^0-E_N'(-\vec{P}+\vec{q})+i\epsilon}\right)
\end{eqnarray}
Then we easily find using Cauchy's theorem that 
\begin{widetext}
\begin{align}
 &-it^{(a)}_{ij}=g_{\Lambda^*,K^-p}\,g_{\Lambda^*,K^-p}\,g_d\frac{D-F}{2f}\int \frac{d^3q}{(2\pi)^3}V_{ij}(q)F(P^0,P^{'0},\vec{q},\omega_K(\vec{q}),\vec{P},\vec{k}) \mathcal{F}^2 (\Lambda, m_K)\nonumber\\
 &-it^{(b)}_{ij}=-g_{\Lambda^*,K^-p}\,g_{\Lambda^*,\pi^+\Sigma^-}g_d\frac{f_{\pi NN}}{m_\pi}\int \frac{d^3q}{(2\pi)^3}W_{ij}(q)F(P^{'0},P^0,\vec{q},\omega_\pi(\vec{q}),-\vec{P},\vec{k})\mathcal{F}^2 (\Lambda, m_\pi)\nonumber\\
 \label{10.1}
 \end{align}
\end{widetext}

where $\mathcal{F} (\Lambda, m_i)$ and $F(P^0,P'^0,\vec{q},\omega,\vec{P},\vec{k})$ are given by
\begin{equation}
 \mathcal{F} (\Lambda, m_i) = \frac{\Lambda^2-m_i^2}{\Lambda^2+\vec{q}^2}  ,
\end{equation}

and

\begin{widetext}
\begin{align}
 &F(P^0,P'^0,\vec{q},\omega,\vec{P},\vec{k})=\frac{1}{2\omega(\vec{q})}\frac{M_N}{E_N(\vec{P}-\vec{q}-\vec{k})}\frac{M_N}{E_N(-\vec{P}+\vec{q})}\frac{M_{\Lambda^*}}{E_{\Lambda^*}(\vec{P}-\vec{q})}\nonumber\\&\times\frac{\theta(q_\mathrm{max}-|\vec{P}-\vec{q}-\frac{\vec{k}}{2}|)}{\sqrt{s}-k^0-E_N(-\vec{P}+\vec{q})-E_N(\vec{P}-\vec{q}-\vec{k})+i\eps}\nonumber\\&\times\left\{\frac{1}{P^0-\omega(\vec{q})-E_{\Lambda^*}(\vec{P}-\vec{q})+i\frac{\Gamma_{\Lambda^*}}{2}}\frac{1}{P^0-\omega(\vec{q})-k^0-E_N(\vec{P}-\vec{q}-\vec{k})+i\eps}\right.\nonumber\\&+\frac{1}{P^0-E_{\Lambda^*}(\vec{P}-\vec{q})-\omega(\vec{q})+i\frac{\Gamma_{\Lambda^*}}{2}}\frac{1}{\sqrt{s}-E_{\Lambda^*}(\vec{P}-\vec{q})-E_N(-\vec{P}+\vec{q})+i\frac{\Gamma_{\Lambda^*}}{2}}\nonumber\\&+\frac{1}{P'^0-E_N(-\vec{P}+\vec{q})-\omega(\vec{q})+i\epsilon}\left.\frac{1}{\sqrt{s}-E_{\Lambda^*}(\vec{P}-\vec{q})-E_N(-\vec{P}+\vec{q})+i\frac{\Gamma_{\Lambda^*}}{2}}\right\}
 \label{10.2}
\end{align}\end{widetext}
and we have introduced a form factor to account for the pseudoscalar exchange with $\Lambda=1125$ MeV as also used in \cite{ikenoraquel}. The functions $V_{ij}(q)$, $W_{ij}(q)$ are the matrix elements of $\vec{\sigma}_2\cdot \vec{q}$ and $\vec{\sigma}_1\cdot \vec{q}$ respectively for the spin transitions $i\to j$ with $i=\uparrow\uparrow, \uparrow\downarrow,\downarrow\uparrow,\downarrow\downarrow$ for the $p\Sigma^-$ spins and $j=\uparrow\uparrow,\frac{1}{\sqrt{2}}(\uparrow\downarrow+\downarrow\uparrow)$, $\downarrow\downarrow$ for the deuteron polarizations. The explicit expressions of $V_{ij}$, $W_{ij}$ are shown in Appendix A. 

The cross section for the $K^-d\to p\Sigma^-$, which is what would be measured, is given by
\begin{equation}
 \frac{d\sigma}{d\mathrm{cos}\,\theta_p}=\frac{1}{4\pi}\frac{1}{s}M_pM_{\Sigma^-}M_d\frac{p}{k}\bar{\sum}\sum \vert t\vert^2\label{eq:dsigco}
\end{equation}
where 
\begin{equation}
 \bar{\sum}\sum \vert t\vert^2=\frac{1}{3}\sum_{i,j}\vert t_{ij}^{(a)}+t_{ij}^{(b)}\vert^2 .
 \label{eq:t2}
\end{equation}
For the evaluation of the $p\Sigma^-\to K^-d$ amplitudes we take $\vec{P}=P(0,0,1)$, $\vec{k}=k(\mathrm{sin}\,\theta_K,0,\mathrm{cos}\,\theta_K)$ and $\vec{q}\equiv q(\mathrm{sin}\,\theta\mathrm{cos}\,\phi, \mathrm{sin}\,\theta\,\mathrm{sin}\,\phi,\mathrm{cos}\,\theta)$. Note that $\mathrm{cos}\,\theta_p$ in $K^-d\to p\Sigma^-$ is the same as $\mathrm{cos}\,\theta_K$ in $p\Sigma^-\to K^-d$.

The factor $p/k$ of phase space makes the cross section blow up as $k\to 0$. For this reason we find appropiate to plot $\frac{k}{p}\left(\frac{d\sigma}{d\mathrm{cos}\theta_p}\right)$, or $\frac{k}{p}\sigma$, after integration over angles.

Since we have two $\Lambda(1405)$ poles, we must sum over them in the $t^{(a)}_{ij}$ or $t^{(b)}_{ij}$ amplitudes. We obtain these amplitudes simply changing the couplings of the resonances to the $K^-p$ or $\pi^+\Sigma^-$ in Eqs.~(\ref{10.1}) and the mass and width of the $\Lambda(1405)$ in Eq.~(\ref{10.2}). As a reference we will use the model of Ref. \cite{40} and we show the properties of these resonances in Table~\ref{tab:poles_OR}. In particular we have
\begin{eqnarray}
 &&g_{\Lambda^*,K^-p}=\frac{1}{\sqrt{2}}g_{\Lambda^*,\bar{K}N}\nonumber\\&&
 g_{\Lambda^*,\pi^+\Sigma^-}=-\frac{1}{\sqrt{3}}g_{\Lambda^*,\pi \Sigma}
\end{eqnarray}
\begin{table}[htb]
\begin{center}
{\renewcommand{\arraystretch}{1}
\setlength\tabcolsep{0.2cm}
 \begin{tabular}{|c|ccc|}\hline
  State&$g_{\Lambda^*,\bar{K}N}$&$g_{\Lambda^*,\pi \Sigma}$&(Mass, $\frac{\Gamma}{2}$)\\\hline
  $\Lambda(1390)$&$1.2+i\,1.7$&$-2.5-i\,1.5$&$(1390,66)$\\
  $\Lambda(1426)$&$-2.5+i\,0.94$&$0.42-i\,1.4$&$(1426,16)$\\\hline
 \end{tabular}}
\end{center}
\caption{Pole positions and couplings from Ref. \cite{47M}.}
\label{tab:poles_OR}
\end{table}

\subsection{Relation to the explicit deuteron wave function}
Following \cite{46} and Appendix A of Ref. \cite{ikenoraquel} (see also Eq. (34) of \cite{ikenoraquel}), one can identify the deuteron wave function is momentum space in our formalism and replace it by the deuteron wave function of the Bonn model \cite{40} (the results with the Paris wave function \cite{paris} are practically the same). The equivalence in the present case is
\begin{widetext}
 \begin{align}
  g_d\frac{M_N}{E(\vec{P}-\vec{q}-\vec{k})}\frac{M_N}{E_N(-\vec{P}+\vec{q})}\frac{\theta(q_{\mathrm{max}}-\vert \vec{P}-\vec{q}-\frac{\vec{k}}{2}\vert)}{\sqrt{s}-k^0-E_N(-\vec{P}+\vec{q})-E_N(\vec{P}-\vec{q}-\vec{k})+i\epsilon}\longrightarrow -(2\pi)^{3/2}\psi(\vec{P}-\vec{q}-\frac{\vec{k}}{2})\label{11.1}
 \end{align}
\end{widetext}
with $\psi(q)$ normalized as $\int d^3 q\vert \psi(\vec{q})\vert^2=1$.
\subsection{Amplitudes using the explicit $\bar{K}N\to \bar{K}N$ and $\bar{K}N\to \pi\Sigma$}
Since formally we have the equivalence
\begin{widetext}
\begin{align}
 &\sum_{i=1}^2\frac{M_{\Lambda^*}^{(i)}}{E_{\Lambda^*}^{(i)}(\vec{P}-\vec{q})}\frac{g^{(i)}_{\Lambda^*,K^-p}g^{(i)}_{\Lambda^*,K^-p}}{\sqrt{s}-E_N(-\vec{p}+\vec{q})-E^{(i)}_{\Lambda^*}(\vec{P}-\vec{q})+i\frac{\Gamma^{(i)}_{\Lambda^*}}{2}}\equiv t_{K^-p,K^-p}(M_\mathrm{inv})\nonumber\\&
 \sum_{i=1}^2\frac{M_{\Lambda^*}^{(i)}}{E_{\Lambda^*}^{(i)}(\vec{P}-\vec{q})}\frac{g^{(i)}_{\Lambda^*,K^-p}g^{(i)}_{\Lambda^*,\pi^+\Sigma^-}}{\sqrt{s}-E_N(\vec{p}+\vec{q})-E^{(i)}_{\Lambda^*}(-\vec{P}-\vec{q})+i\frac{\Gamma^{(i)}_{\Lambda^*}}{2}}\equiv t_{K^-p,\pi^+\Sigma^-}(M'_\mathrm{inv})
\end{align}
\end{widetext}
with $M_{\mathrm{inv}}^2=s+M^2_N-2\sqrt{s}E_N(-\vec{P}+\vec{q}) $, and $M_{\mathrm{inv}}'^2=s+M^2_N-2\sqrt{s}E_N(\vec{P}+\vec{q})$, respectively. We can write the amplitudes $t_{ij}^{(a)}$, $t_{ij}^{(b)}$ as
\begin{widetext}
\begin{align}
 &-it^{(a)}_{ij}=g_d\frac{D-F}{2f}\int \frac{d^3q}{(2\pi)^3}V_{ij}(q)F'(P^0,P'^{0},\vec{q},\omega_K(\vec{q}),\vec{P},\vec{k}) \mathcal{F}^2 (\Lambda, m_K)\nonumber\\
 &-it^{(b)}_{ij}=-g_d\frac{D+F}{2f}\int \frac{d^3q}{(2\pi)^3}W_{ij}(q)G'(P^{0},P'^0,\vec{q},\omega_\pi(\vec{q}),\vec{P},\vec{k})\mathcal{F}^2 (\Lambda, m_\pi)\nonumber\\
 \label{12.2}
 \end{align}
 \end{widetext}
where $F'$ and $G'$ are given in Appendix B.
                      
\section{Results}
\begin{figure}
 \centering
 \includegraphics[width=0.45\textwidth]{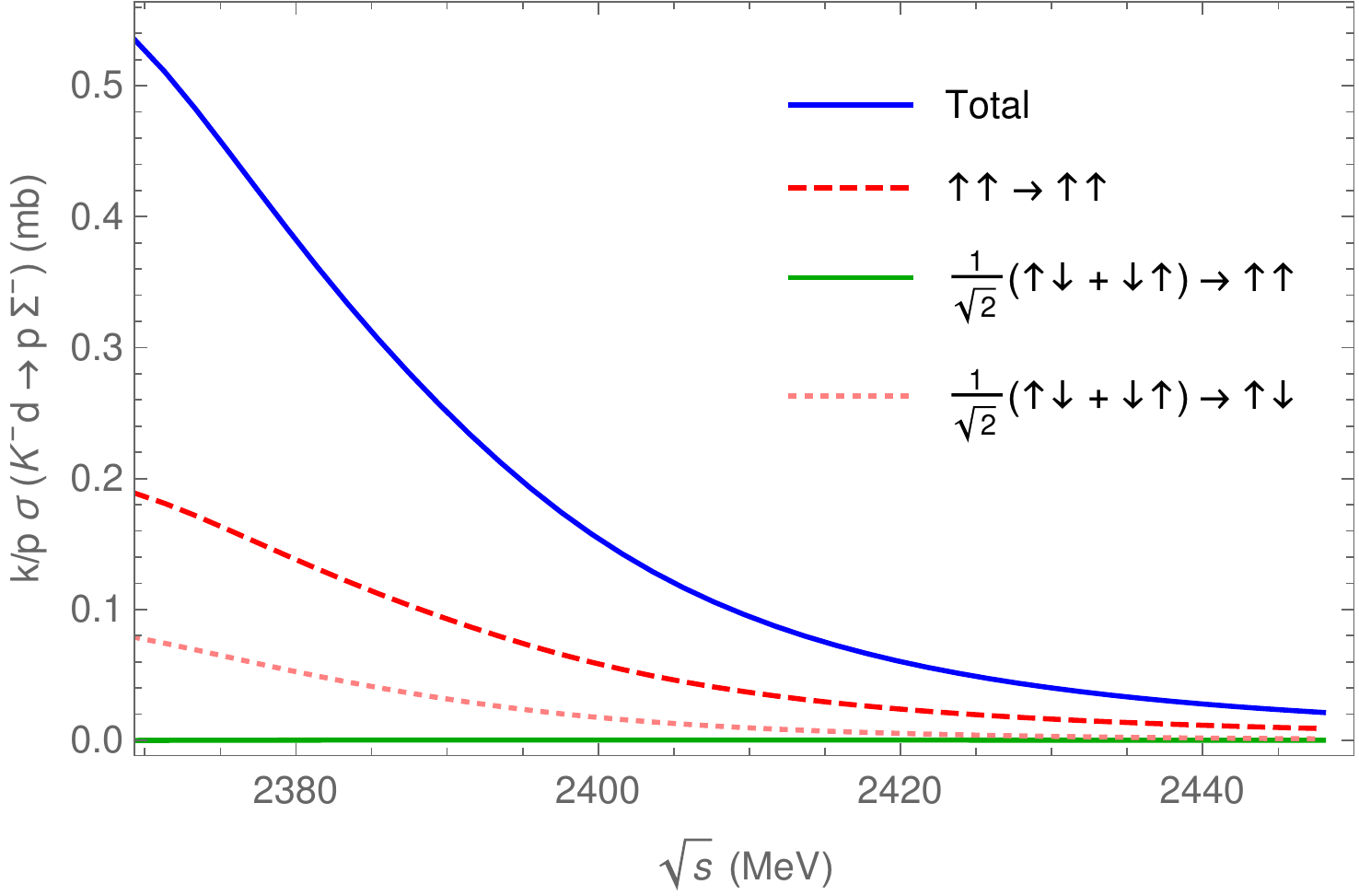}
 \caption{Contribution from several spin transitions in $\frac{k}{p}\sigma$.}
 \label{fig:figure_3}
\end{figure}

\begin{figure}
 \centering
 \includegraphics[width=0.45\textwidth]{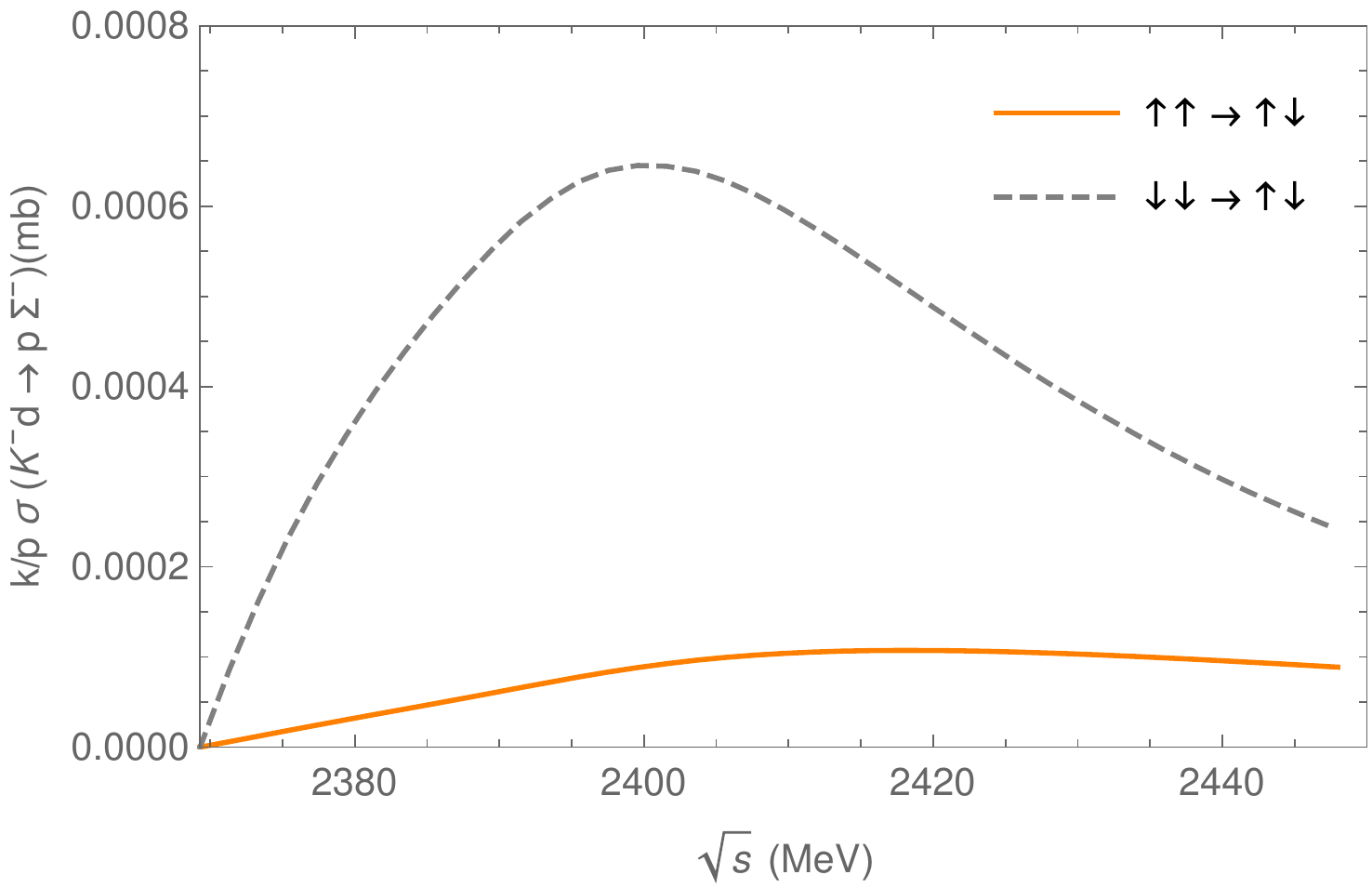}
 \caption{Contribution from several spin transitions in $\frac{k}{p}\sigma$.}
 \label{fig:figure_4}
\end{figure}

In the first place, we study the contribution of the different spin transitions. We use the model of Ref.~\cite{40M} (called Oset-Ramos later) taking the input of Table~\ref{tab:poles_OR}. In Fig.~\ref{fig:figure_12}, we shall present results for the cross sections using different models. The cross sections are taken from Eq.~(\ref{eq:dsigco}) subsequently integrated over the angle $\theta_p$. In Fig.~\ref{fig:figure_3} we plot $\frac{k}{p}\sigma$ for several spin transitions among which one finds that the most important is $\uparrow\uparrow \to \uparrow\uparrow$, or $\downarrow\downarrow \to \downarrow\downarrow$ which has the same strength. The transitions involving some spin flip are very small as shown in Fig.~\ref{fig:figure_4}. This spin dependence is different from the one obtained in the 
$pp \to \pi^+d$ reaction \cite{ikenoraquel} and the reason is that, unlike in \cite{ikenoraquel}, we have only a spin operator in one of the baryonic lines.\\

\begin{figure*}
 \centering
 {\renewcommand{\arraystretch}{1}
\setlength\tabcolsep{0.1cm}
 \begin{tabular}{lc}
\hspace{-0.5cm} \includegraphics[width=0.45\textwidth]{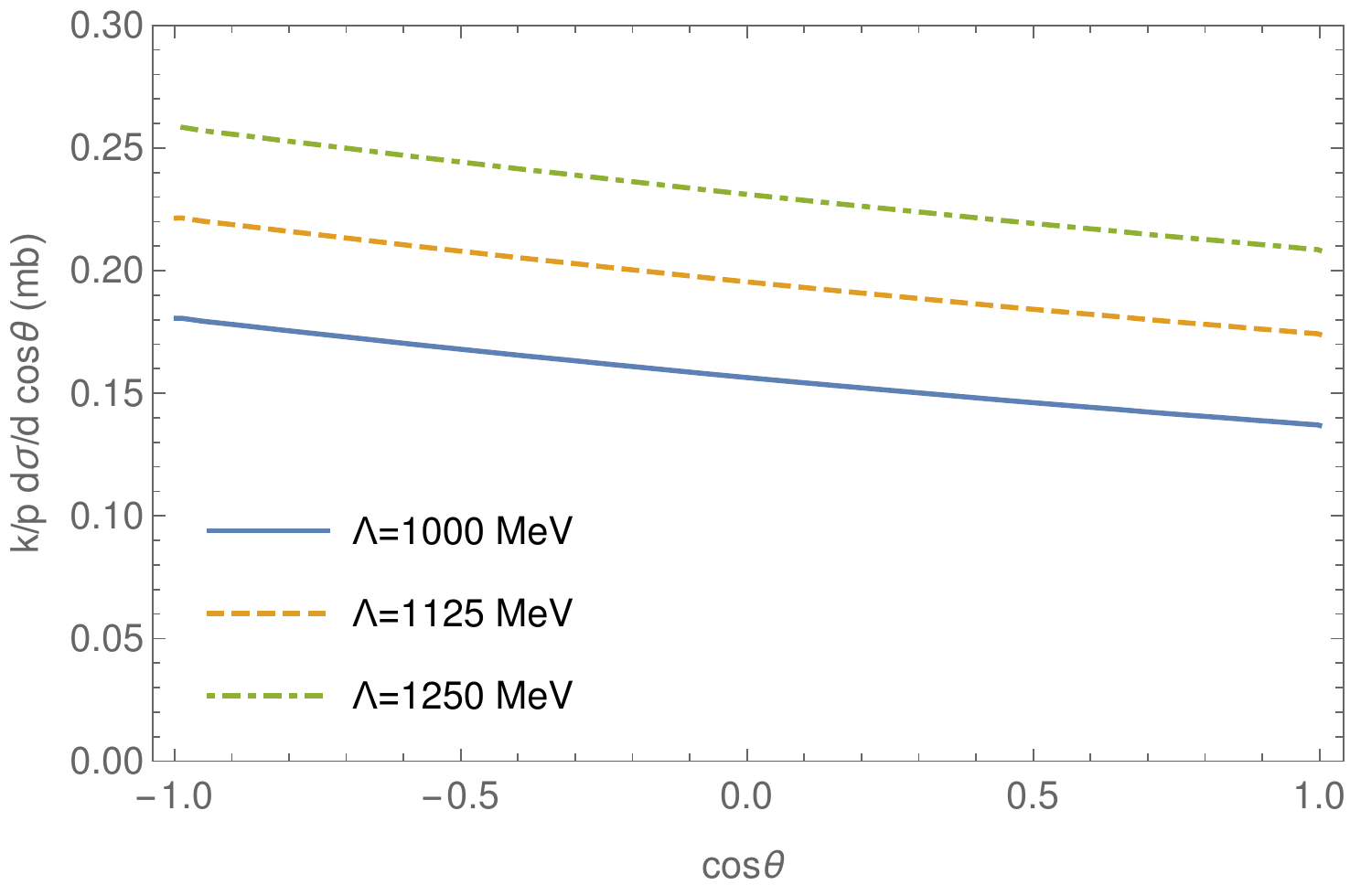}&\includegraphics[width=0.45\textwidth]{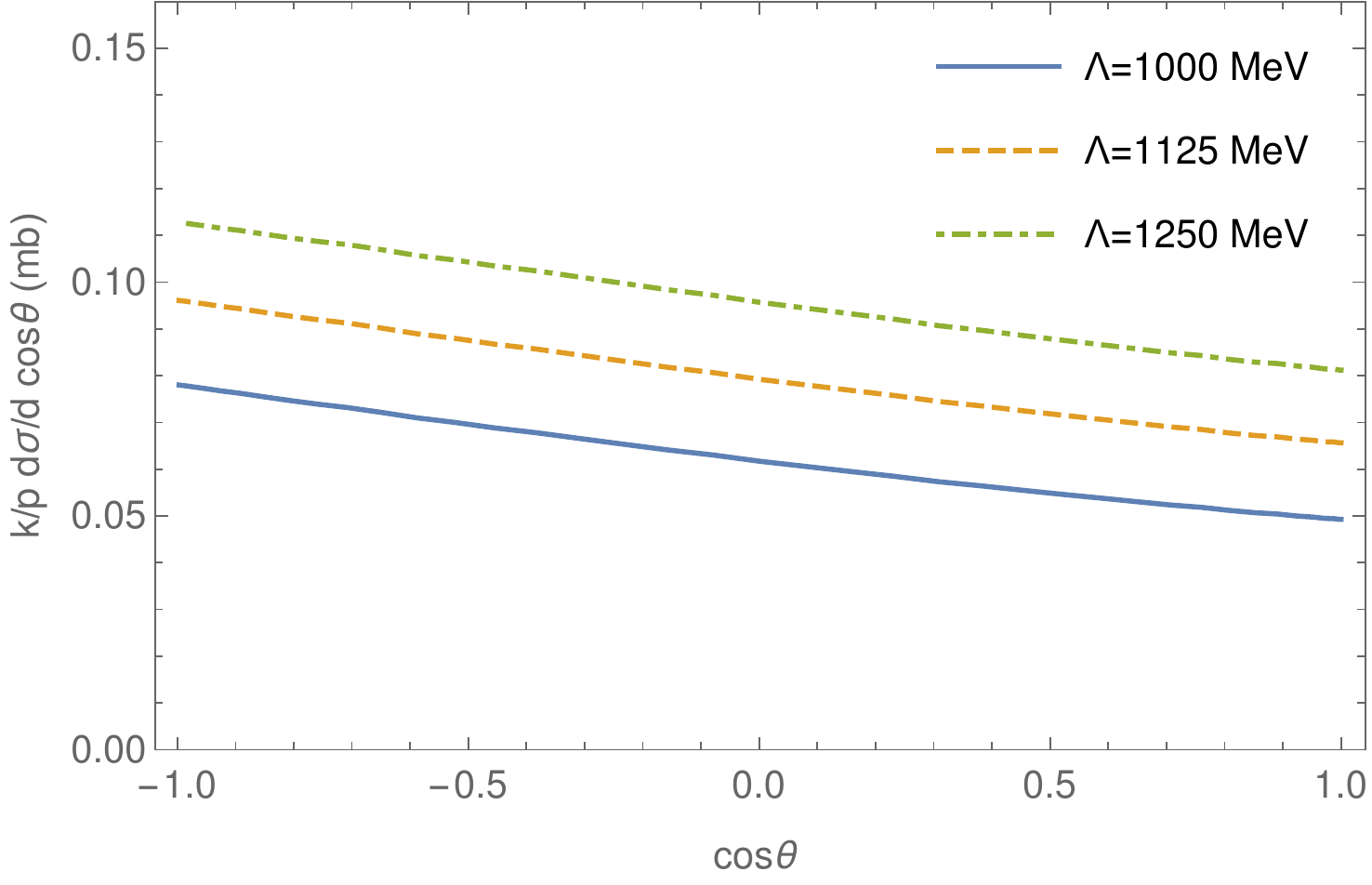}\\
 \end{tabular}}
 \caption{Angular dependence of $\frac{k}{p}\frac{d\sigma}{d\mathrm{cos}\,\theta_p}$ of the $K^- d \to p \Sigma^-$ reaction as a function of $\mathrm{cos}\theta$ for $\sqrt{s}=\sqrt{s}_{\mathrm{th}}+10$ MeV (left) and $\sqrt{s}_{\mathrm{th}}+30$ MeV (right).}
 \label{fig:figure_5}
\end{figure*}

Next, we look at the angular dependence. Fig.~\ref{fig:figure_5} displays the angular dependence of $d\sigma/d\mathrm{cos}(\theta_p)$ for two given values of $\sqrt{s}$, namely about $10$~MeV and $30$~MeV above threshold.

We observe a smooth angular dependence, a little stronger as the energy increases, favoring backward angles.

Even if the lines look parallel, there are small differences in the slope for different values of $\Lambda$. Indeed, in Fig.~\ref{fig:figure_5} (left), the ratio of backward to forward cross section is a factor $1.24$ for $\Lambda=1250$~MeV, while it is $1.32$ for $1000$ MeV. The differences are bigger in Fig.~\ref{fig:figure_5} (right), at higher energies, where these ratios are $1.42$ and $1.58$ respectively. A precise measurement of the angular distributions can tell us about the value of $\Lambda$ to be used. We should also note that a difference of $125$~MeV in the value of $\Lambda$ induces a change in the cross section of $16-18$\% around threshold. We shall see comparing models that the differences in the cross sections predicted by different models are of the order of a factor two. Hence, even with small uncertainties in the value of $\Lambda$, the measured cross sections can serve to discriminate among models. \\

\begin{figure*}
 \centering
 {\renewcommand{\arraystretch}{1}
\setlength\tabcolsep{0.1cm}
 \begin{tabular}{lc}
\hspace{-0.5cm} \includegraphics[width=0.45\textwidth]{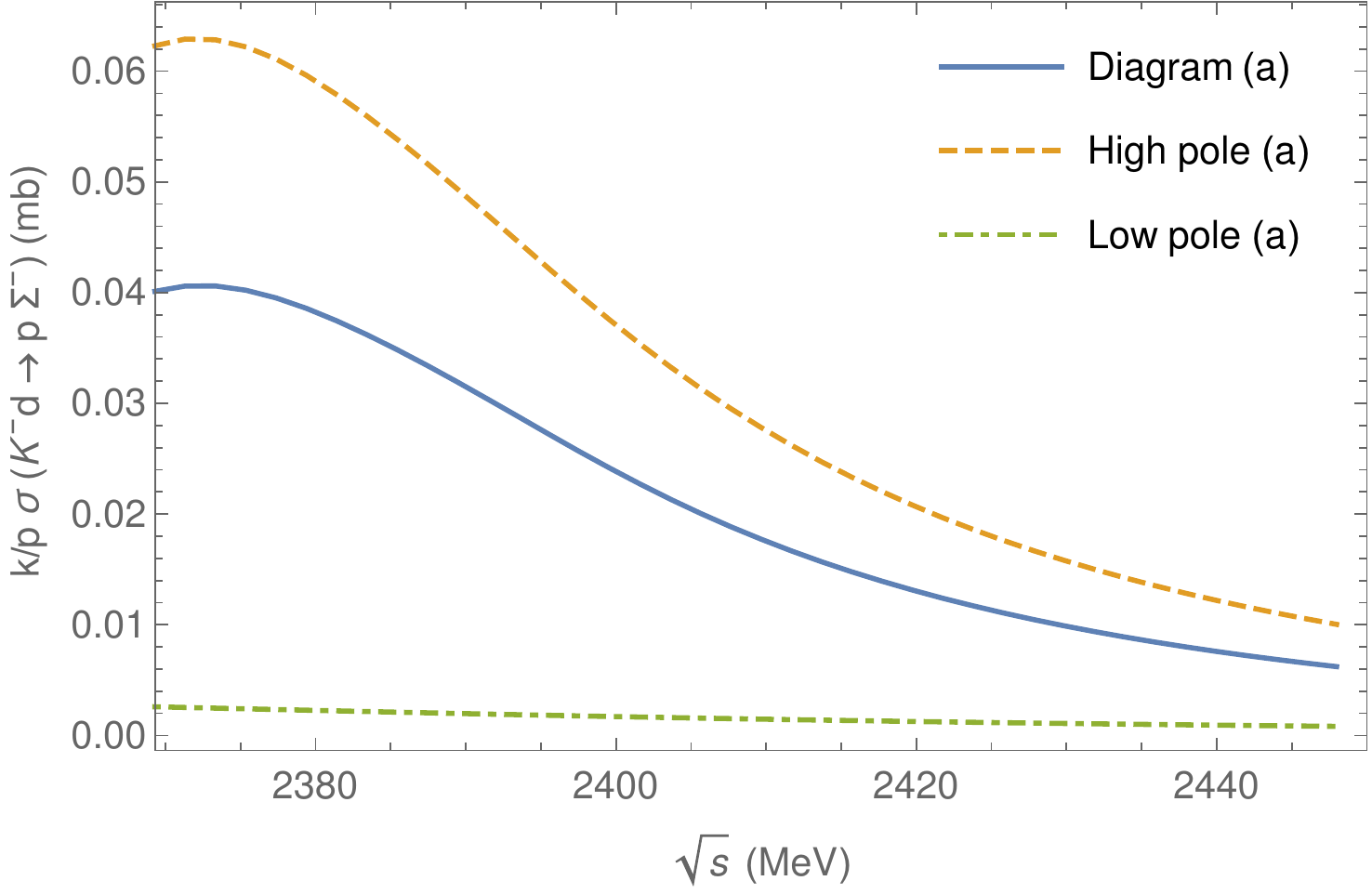}&\includegraphics[width=0.45\textwidth]{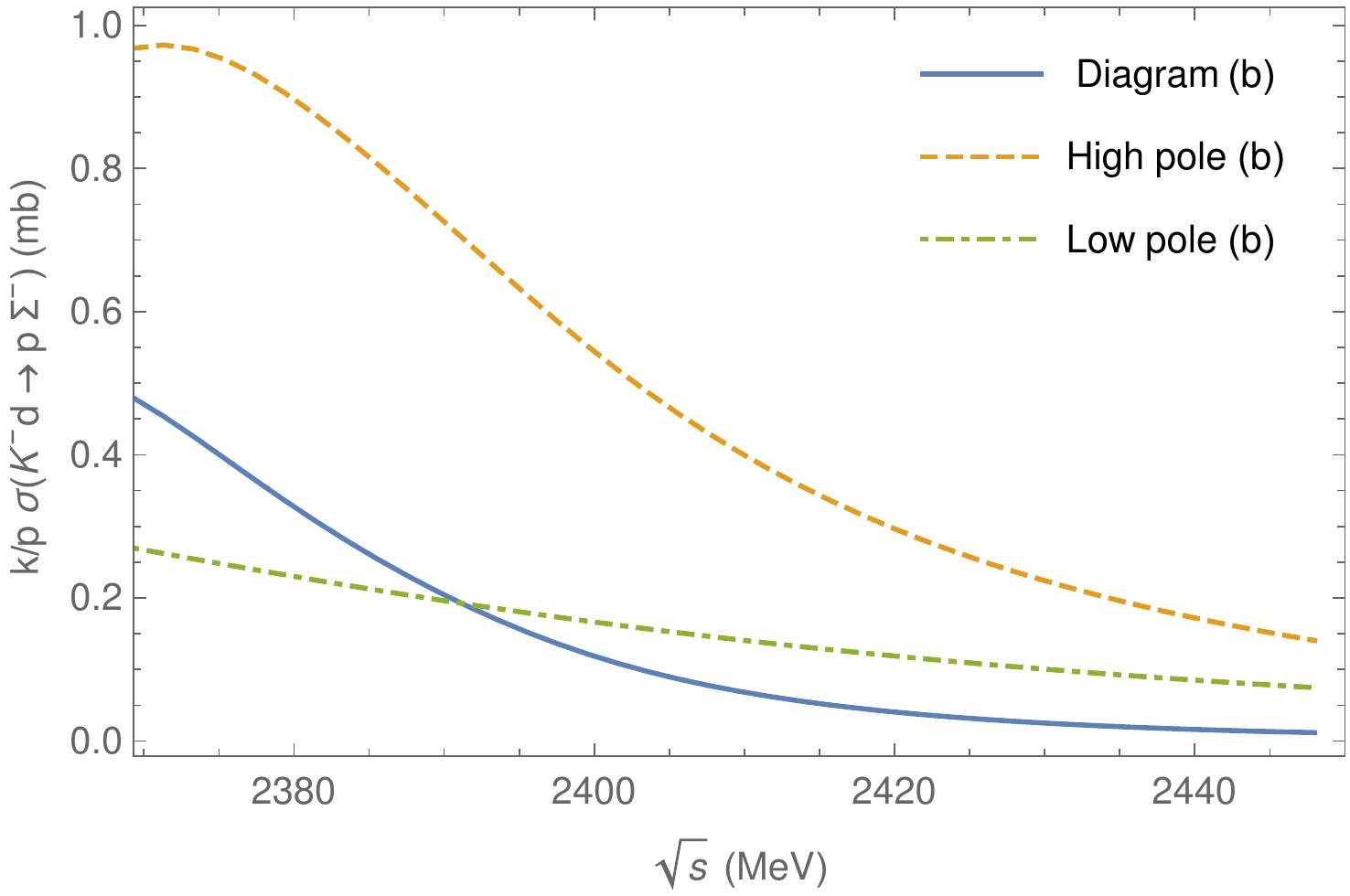}\\
 \end{tabular}}
 \caption{Contributions of the high and low mass poles to the diagrams (a) and (b) in $\frac{k}{p}\sigma$.}
 \label{fig:figure_6}
\end{figure*}

In Fig.~\ref{fig:figure_6} (left), we show the contribution of each of the $\Lambda(1405)$ poles for the mechanism of Fig.~\ref{fig:2} (a). As one might expect from Eq.~(\ref{7.1}), we observe that the contribution of the higher mass pole is much larger than that of the lower mass pole. However, the destructive interference between the two contributions is relevant enough to reduce the cross section in about $30$\%.

An analogous study is performed for the mechanism diagrammatically represented in Fig.~\ref{fig:2} (b). Fig.~\ref{fig:figure_6} (right) shows qualitatively similar features to the former case, but with approximately a factor ten difference in the strength of the cross section between both mechanisms. The reason for this lies in the magnifying effect of the pseudoscalar propagator  produced by the fact of having a pion exchange instead of a kaon exchange. We also note that now the interference is more apparent and reduces the cross section by about $50$\%. \\

\begin{figure*}
 \centering
 {\renewcommand{\arraystretch}{1}
\setlength\tabcolsep{0.1cm}
 \begin{tabular}{lc}
\hspace{-0.5cm} \includegraphics[width=0.45\textwidth]{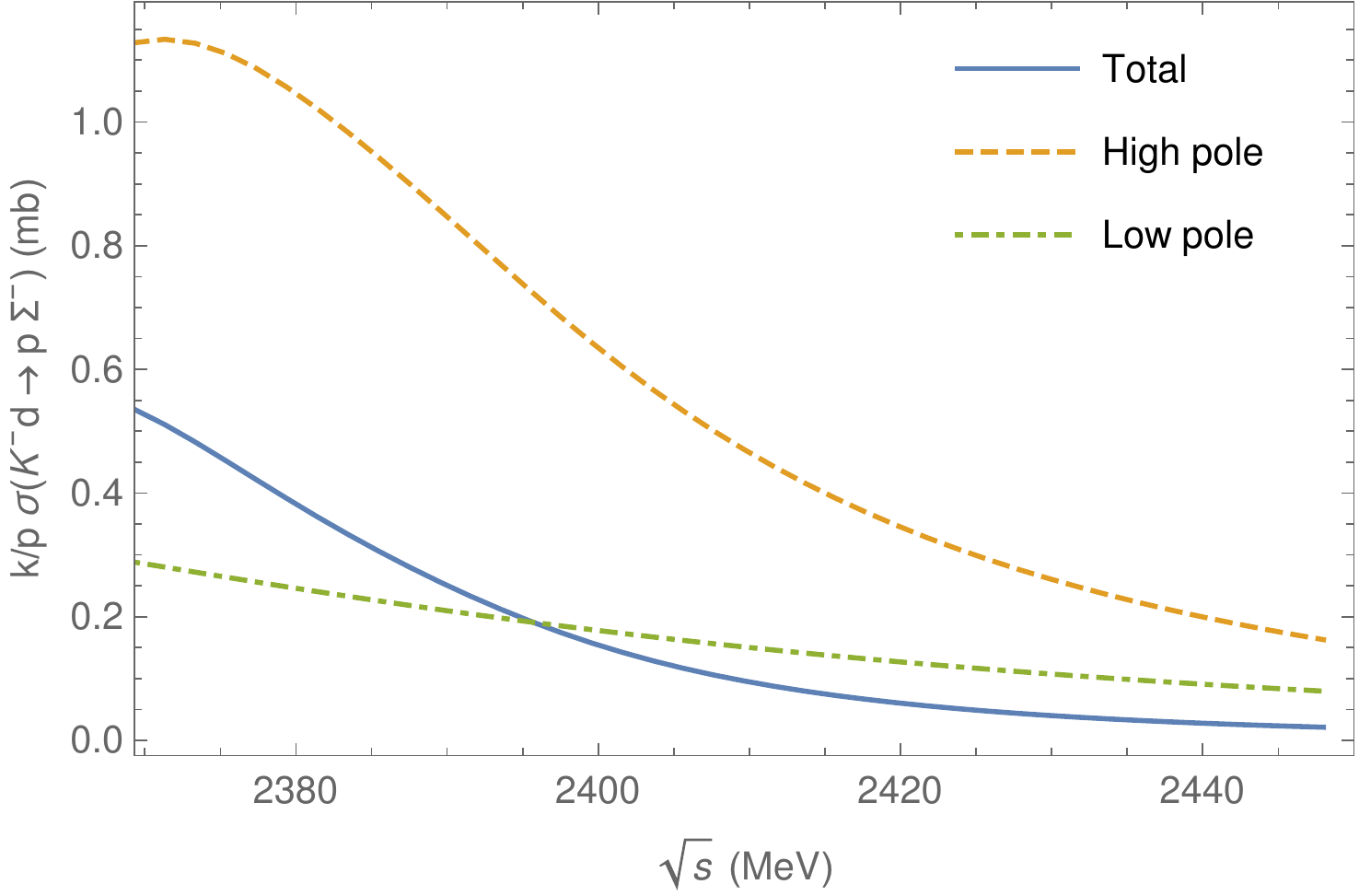}&\includegraphics[width=0.45\textwidth]{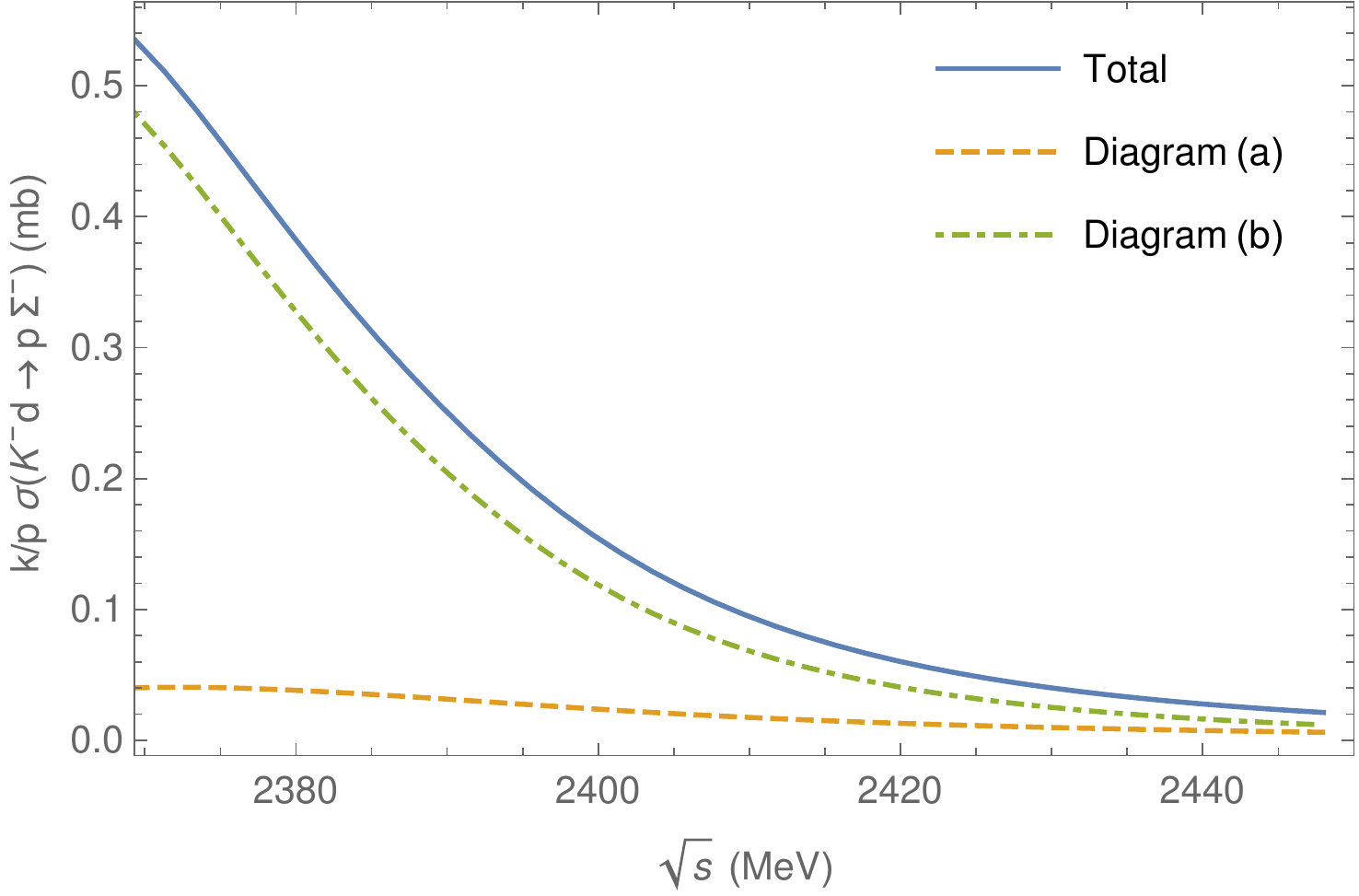}\\
 \end{tabular}}
 \caption{Contributions of the high and low mass poles (left) and diagrams (a) and (b) (right), compared to the total $\frac{k}{p}\sigma$.}
 \label{fig:figure_8}
\end{figure*}

In Fig.~\ref{fig:figure_8} (left), we show the results for each of the poles once the contributions of the two mechanisms are included in the cross section simultaneously. As before, we find the dominance of the higher $\Lambda(1405)$ pole, but the interference reduces the cross section to one half its strength.\\

The role of each mechanism in the $K^- d \to p \Sigma^-$ process is reflected in Fig.~\ref{fig:figure_8} (right) including the contribution of both poles. The results when independently taking the contributions of the two mechanisms confirm that the mechanism of Fig. \ref{fig:2} (b) is the dominant one. As an interesting fact here, it should be mentioned that the addition of the mechanism of Fig.~\ref{fig:2} (a) to the one of Fig. \ref{fig:2} (b) increases the cross section by just $10$\%.\\

\begin{figure}
 \centering
 \includegraphics[width=0.45\textwidth]{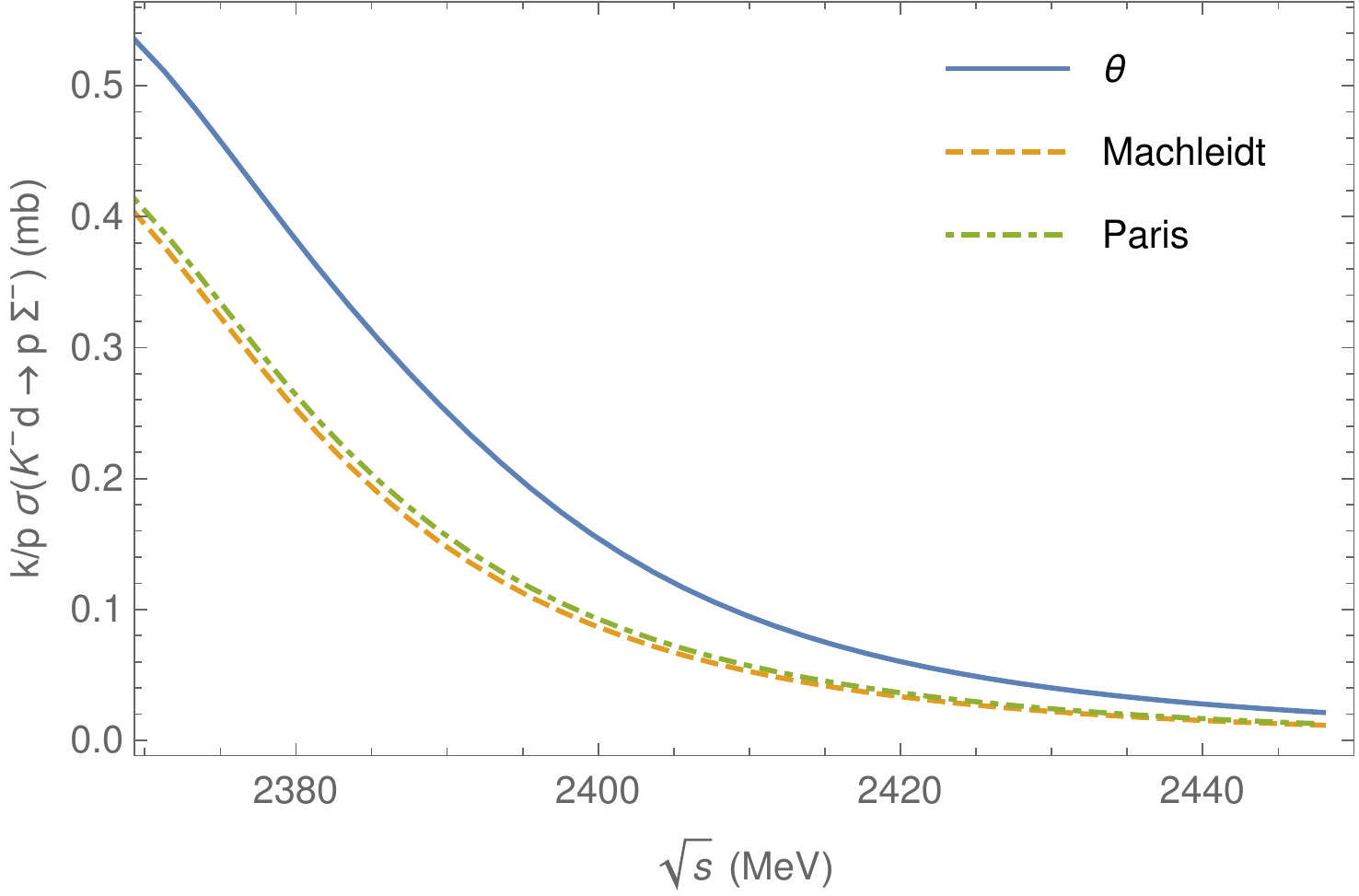}
 \caption{Comparison of $\frac{k}{p}\sigma$ using the $\theta$ function or the deuteron wave funtions.}
 \label{fig:figure_9}
\end{figure}

So far all the results shown have been obtained employing the $\theta$-function as prescription for the deuteron wave function. Next, using Eq.~(\ref{11.1}), we swap the former prescription for the explicit wave functions derived from the Bonn and Paris potentials \cite{38,paris}. The results are collected in Fig.~\ref{fig:figure_9}. We observe that the cross sections with either wave function are very similar, but with respect to the $\theta$ wave function they reduce the cross section by about $23$\%. An attempt to reduce the previous difference was carried out by rescaling the $g_d$ coupling by a factor $0.88$ that works fairly well with small values of $\sqrt{s}$ but does not mantain such an accuracy as the energy increases. Because of this, from now on all the results presented are obtained from the Bonn wave function. \\

\begin{figure*}
 \centering
 {\renewcommand{\arraystretch}{1}
\setlength\tabcolsep{0.1cm}

 \begin{tabular}{lc}
\hspace{-0.5cm} \includegraphics[width=0.45\textwidth]{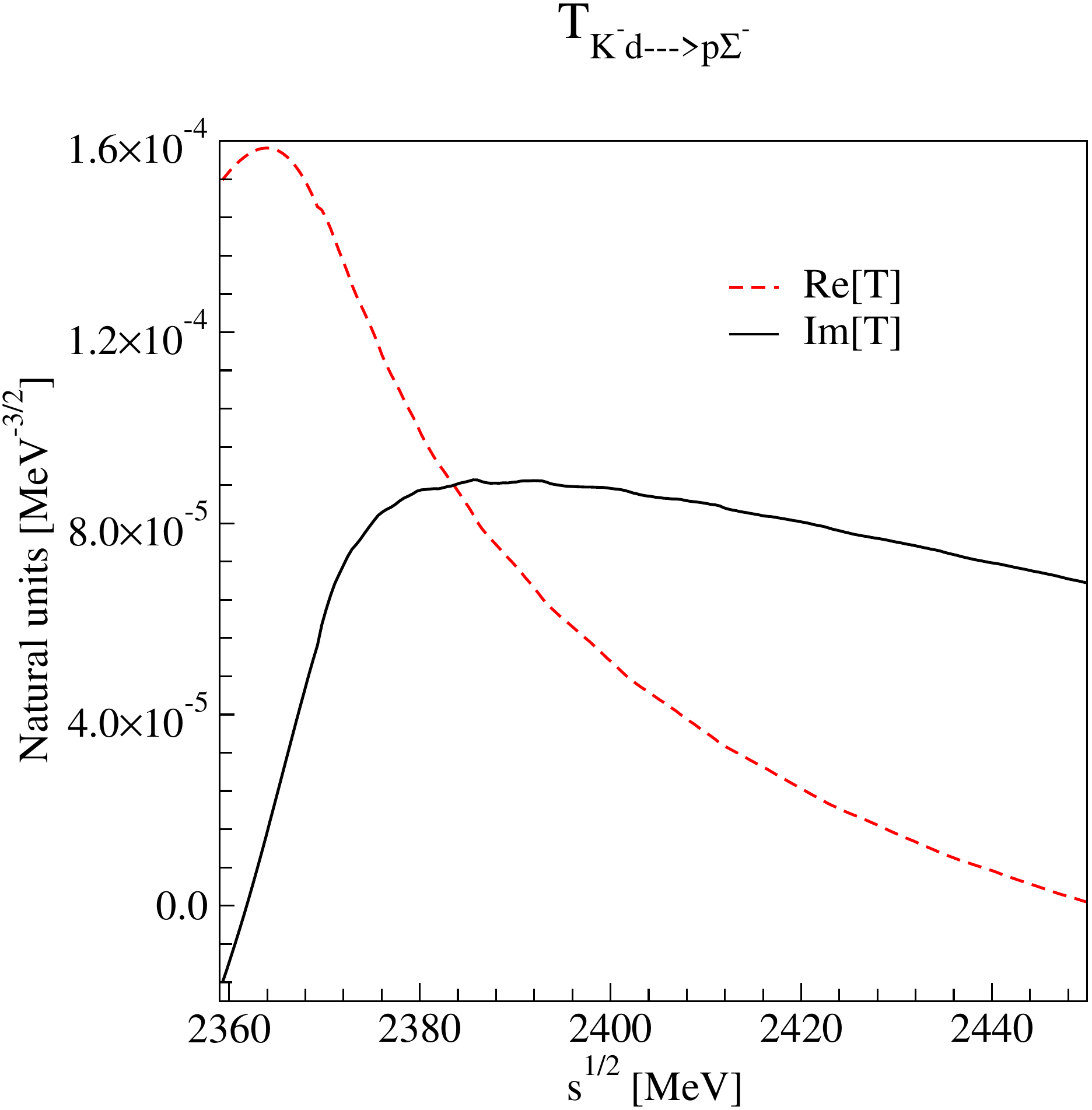}&\includegraphics[width=0.45\textwidth]{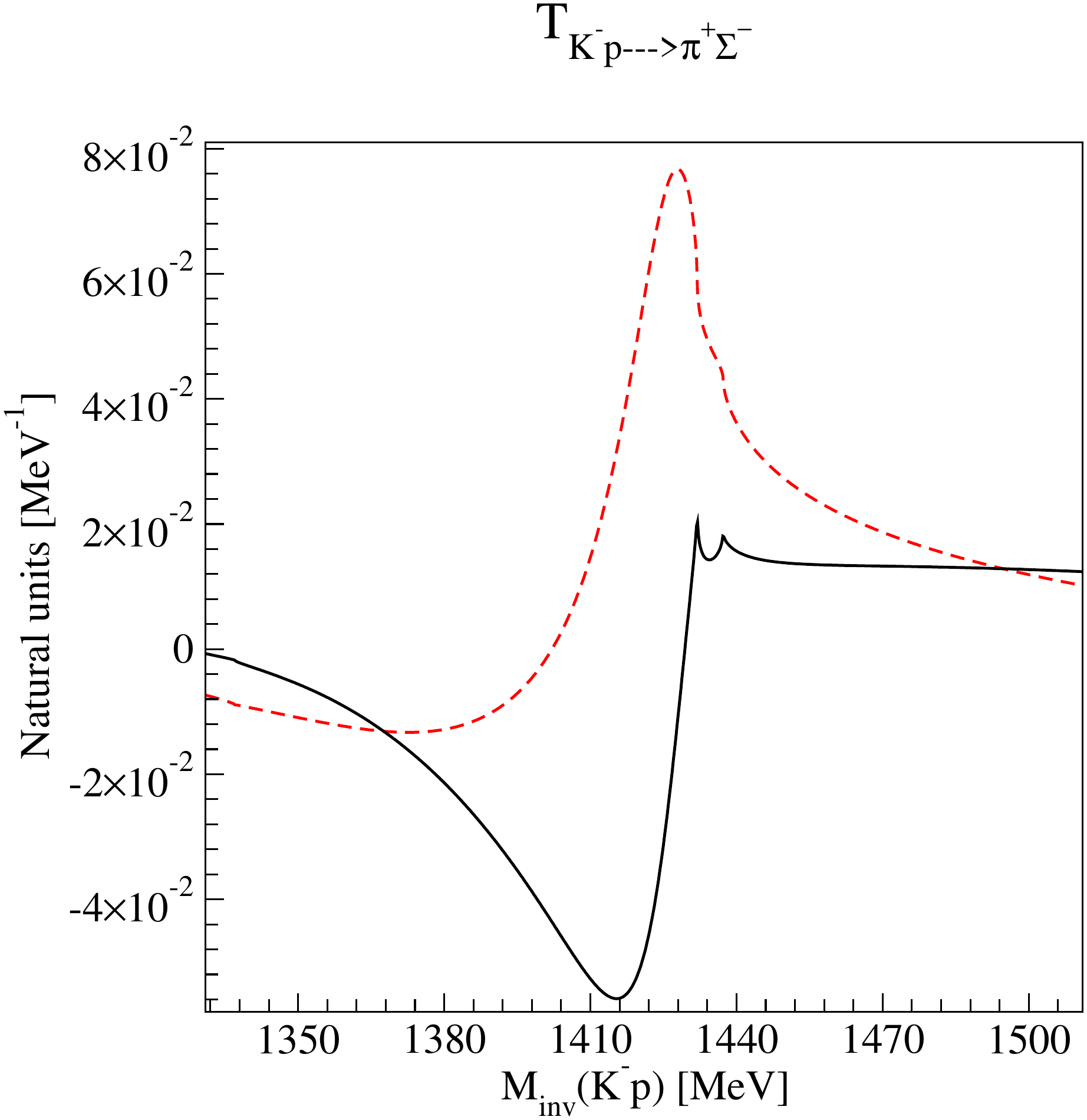}\\
 \end{tabular}}

 \caption{Energy dependence of the real (dashed line) and the imaginary (solid line) parts of the $K^- d \to p \Sigma^-$ (left panel)  and $K^- p \to \pi^+ \Sigma^-$ (right panel) amplitudes.}\label{fig:figure_11}
\end{figure*}

\begin{figure*}
 \centering
 {\renewcommand{\arraystretch}{1}
\setlength\tabcolsep{0.1cm}
\includegraphics[width=0.65\linewidth]{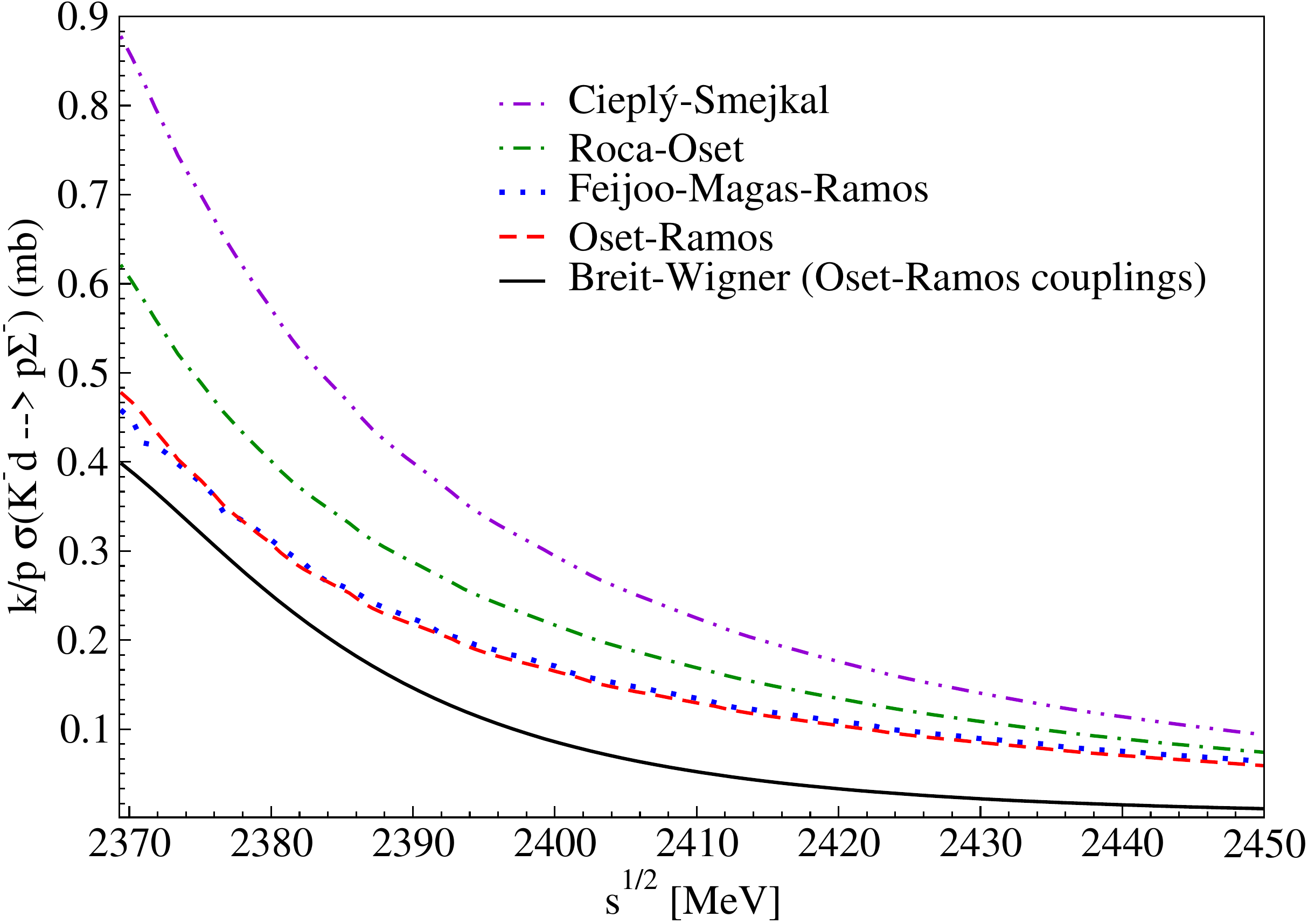}}
 \caption{$K^- d \to p \Sigma^-$ cross sections ($\frac{k}{p}\sigma$) for the considered models. More details in the text.}
 \label{fig:figure_12}
\end{figure*}

In order to illustrate the effect of the triangle singularity on the $K^- d \to p \Sigma^-$ reaction, we find very instructive to show the corresponding amplitude. For simplicity, we have chosen the dominant spin transition $\uparrow\uparrow \to \uparrow\uparrow$ and plot the real and imaginary parts of its amplitude in the left panel of Fig.~\ref{fig:figure_11}. 
Both curves behave as one might expect for a typical triangle singularity. On the one hand, we see that the imaginary part (solid line) develops a smooth peak, with its highest strength around $\sqrt{s}=2380$ MeV, close to where Eq.~(\ref{7.1}) predicts the peak of the TS. The shape of the imaginary part translates information of the imaginary part of the $t_{K^- p, \pi^+ \Sigma^-}$ amplitude below threshold to $K^-$ energies above threshold in the $K^- d \to p \Sigma^-$ reaction. On the other hand,  the real part of the amplitude contributes largely to the strength of the cross section close to threshold. The dashed line in the same panel of Fig.~\ref{fig:figure_11} clearly evidences the presence of a cusp in the vicinity of the $K^- d$ threshold. To aid the visualization of this effect, we have explored a bit below threshold setting $\vec{k}=0$ in the formulas. These findings are in good agreement with Ref.~\cite{sacairamos},  where it was shown that the imaginary part of the amplitude is tied to the triangle singularity, while the real part is tied to a threshold effect. 
In the right panel of Fig.~\ref{fig:figure_11}, we show the real and imaginary parts of the $t_{K^- p, \pi^+ \Sigma^-}$ amplitude, the one that dominates the reaction in the mechanism of Fig.~\ref{fig:2} (b). As one can see in Eqs.~(\ref{12.2}, \ref{eq:fs}), the shape of the $K^- d \to p \Sigma^-$ amplitude is not a mapping of the one from $t_{K^- p, \pi^+ \Sigma^-}$. Actually, comparing both panels, it can be appreciated that the TS has created a structure of its own thereby making the amplitudes differ significantly from each other. Nevertheless, by construction, Eqs.~(\ref{12.2}, \ref{eq:fs}) constrain the shape and strength of $t_{K^- d , p \Sigma^-}$ to be strongly tied to that of $\bar{K} N \to \pi \Sigma$ amplitude. Moreover, the terms in $F'$ and $G'$ of Eq.~(\ref{eq:fs}) (Apendix B) showing explicitly the couplings of the $\Lambda(1405)$ resonances to the $\bar{K} N$ and $\pi \Sigma$ channels give a small contribution of the order of $5$\% in the $K^- d \to p \Sigma^-$ amplitudes. Therefore, this last magnitude is basically proportional to the $K^- p \to \pi^+ \Sigma^-$ amplitude, except that the amplitude, weighted by some factors changing with the loop integration variables, is integrated for different values of the invariant mass in the loop. However, one should bear in mind that most of the contribution comes from the region of the integration variables where the singularity given by Eq.~(\ref{7.1}) appears. This gives chances for the reaction to show a sensitivity to different models, which we exploit below. \\

Finally, we analyze the impact of the $K^-p \to K^-p, \pi^+ \Sigma^-$ amplitudes from different models on the  $K^- d \to p \Sigma^-$ cross section. To this end, four models have been considered whose nature could be representative of what one can find in the literature. All of them are derived from a chiral $SU(3)$ Lagrangian and implementing a unitarization scheme in coupled channels, as well as limited to s-wave projection. Despite this common approach, there are some peculiarities worth mentioning that can be useful for a future understanding of potential experimental data. 


 {\bf  Oset-Ramos} \cite{40M}: This first model uses a Weinberg-Tomozawa (WT) contribution as driving term in the interaction kernel. The authors took into account, for the first time, the full $S=-1$ meson-baryon basis for the regularization of loop integral in coupled channels.  

 {\bf Roca-Oset} \cite{luisone}: This model takes as building block the previous one, yet reducing the basis to $\pi \Lambda, \pi \Sigma, \bar{K}N$ channels. It has special interest for the present study because, apart from the ordinary $K^- p \to \pi \Lambda,  \pi \Sigma, \bar{K}N$ total cross sections and threshold observables, the authors incorporated the CLAS data for the $\Lambda(1405)$ photoproduction \cite{Moriya:2013eb} in the fits to constrain the model parameters. 

 {\bf Ciepl\'y-Smejkal}: The third one is the model called NLO30 in Ref.~\cite{59M}. This is a model based on a chirally motivated potential, written in a separable form, whose central piece is derived from the Lagrangian up to next-to-leading order (NLO). The authors took into account the very precise measurements of the shift and width of the 1S state in kaonic hydrogen carried out by SIDDHARTA Collaboration in the fitting procedure \cite{Bazzi:2011zj}.\\

 {\bf  Feijoo-Magas-Ramos}: This model corresponds to the fit called WT+Born+NLO carried out in \cite{Feijoo:2018den}. It was constructed by adding the interaction kernels derived from the Lagrangian up to NLO, and including additional experimental data at higher energies in the fits. This model is the natural extension of Oset-Ramos.\\


In Fig.~\ref{fig:figure_12} we plot the resulting $K^- d \to p \Sigma^-$ cross sections for the models discussed above. The most revealing feature in the figure is the significant difference in the cross section strength among them at small values of $\sqrt{s}$. The Oset-Ramos (dashed line) and Feijoo-Magas-Ramos (dotted line) produce almost identical results. The discrepancies become more evident when comparing the former models with Oset-Roca (dash-dotted) and Ciepl\'y-Smejkal (dash-dot-dot line) being of the order of  $40$\%  and  $100$\%, respectively. For low values of $\sqrt{s}$ in the $K^- d \to p \Sigma^-$ reaction, the contributions of the $\bar{K}N$ amplitudes in the loop integral come mostly from $\bar{K}N$ invariant masses in the subthreshold region where the $\bar{K}N$ models present the greatest disagreements. One expects that, as the energy of the  $K^-$ increases, the $\Lambda(1405)$ invariant mass, still restricted by the nucleon dynamics in the deuteron, has more access to values  where all models have a better agreement among themselves because they have been fitted to the same experimental data. This effect is somewhat seen in Fig.~\ref{fig:figure_12} by the convergent trend shown by the models at higher energies. Roughly speaking, the process proposed in the present study acts as an indirect window to the subthreshold $\bar{K}N$ amplitudes that cannot only be used as a tool to discern which models are suitable to describe the physics in such a region but also may shed some light on the location of the lower mass pole of the $\Lambda(1405)$ resonance.\\

\section{Conclusions}

We have investigated the $p \Sigma^- \to K^- d$ and its time reversal $K^- d \to p \Sigma^-$ reactions, which are driven by a triangle mechanism with the $\Lambda(1405)$, a proton and a neutron in the intermediate states. We show that the triangle mechanism develops a triangle singularity which magnifies the cross section and produces a particular shape in the cross section. We show analytically that in the case of a narrow $\Lambda(1405)$ width, a TS appears a few MeV above threshold, and this peak becomes broader upon consideration of the $\Lambda(1405)$ width. We could show that of the mechanisms involving a $\pi$ or $K$ exchange, the one involving the $\pi$ exchange is the dominant one, and of the two $\Lambda(1405)$ resonances, the one of higher mass gives also the largest contribution. We showed, from the analytical expression of the transition amplitude, that it was weighting the $K^- p \to \pi^+ \Sigma^-$ amplitude below threshold with a particular configuration tied to the TS which produced a shape quite distinct from the one of the $K^- p \to \pi^+ \Sigma^-$ amplitude. This dependence on the $\bar{K} N$ and $\pi \Sigma$ amplitude below threshold makes this reaction quite sensitive to different models that, giving similar cross sections for $\bar{K} N$ reactions above threshold, produce rather different extrapolations of the $\bar{K} N$ amplitudes below threshold. This information is relevant in the issue of $\bar{K}$ bound states in nuclei \cite{gal,hirenzaki,baca} and, thus, the measurement of this reaction will provide new and valuable information concerning this problem. On the other hand, concerning the two poles of the $\Lambda(1405)$, one around $1420$ MeV and the other one around $1385$ MeV, while practically all theoretical models coincide on the features of the $\Lambda(1420)$, they differ substantially in the the position and width of the lower mass one. The new information provided by this reaction will help to narrow the predictions around the second state. \\

\section*{Acknowledgments}

We would like to thank Ale\v s Ciepl\'y for providing us with the $\bar{K}N$ scattering amplitudes calculated from NLO30 model as well as the corresponding $\Lambda(1405)$ pole positions and their couplings.
R. M. acknowledges support from the CIDEGENT program with Ref. CIDEGENT/2019/015 and from the spanish national grant PID2019-106080GB-C21. This work is also partly supported by the Spanish Ministerio de Economia y Competitividad and European FEDER funds under Contracts No. FIS2017-84038-C2-1-P B and No.  FIS2017-84038-C2-2-P B. This project has also received funding from the European Union’s Horizon 2020 programme 
No. 824093 for the STRONG-2020 project and by Generalitat Valenciana under contract PROMETEO/2020/023. 
The work of A. F. was partially supported by the Czech Science Foundation, GA\v CR Grant No. 19-19640S. 
L. R. D. acknowledges the support from the National Natural Science Foundation of China (Grant Nos. 11975009, 11575076).

\appendix

\section{Spin matrix elements $V_{ij}$ and $W_{ij}$.}

The $\vec{\sigma}\cdot\vec{q}$ product is expressed as:
\begin{equation}
\vec{\sigma}\cdot\vec{q}=\sigma_+ q_- + \sigma_- q_+ + \sigma_z q_z
\end{equation}
with
\begin{eqnarray}
q_+ &=& q_x + {\rm i\,}q_y \nonumber\\
q_- &=& q_x - {\rm i\,}q_y  \nonumber\\
\sigma_+ &=& \frac{1}{2}(\sigma_x + {\rm i\,}\sigma_y) \nonumber\\
\sigma_- &=& \frac{1}{2}(\sigma_x - {\rm i\,}\sigma_y).
\end{eqnarray}

Then, the elements of the spin-transition matrix for a couple of incoming  $\frac{1}{2}$-baryons ($p$ and $\Sigma^-$, in this case) going to the constituent pair of nucleons merged in the deuteron (S=1) can be written in the following form.

Accounting for {\bf mechanism (a)} in  Eq.~(\ref{10.1})), $V_{ij}$:
\begin{eqnarray}
 \uparrow\uparrow \to \uparrow\uparrow:                                 \;\;\;\;             V_{11}&=&q_z    \nonumber\\
 \uparrow\uparrow \to \frac{1}{\sqrt{2}}(\uparrow\downarrow+\downarrow\uparrow): \;\;\;\;    V_{12}&=&\frac{1}{\sqrt{2}}q_+   \nonumber\\
 \uparrow\uparrow \to \downarrow\downarrow:                    \;\;\;\;                      V_{13}&=&0    \nonumber\\  
 \uparrow\downarrow \to \uparrow\uparrow:                       \;\;\;\;                     V_{21}&=&q_-   \nonumber\\
 \uparrow\downarrow \to \frac{1}{\sqrt{2}}(\uparrow\downarrow+\downarrow\uparrow): \;\;\;\;  V_{22}&=&-\frac{1}{\sqrt{2}}q_z \nonumber\\
 \uparrow\downarrow \to \downarrow\downarrow:      \;\;\;\;                                  V_{23}&=&0    \nonumber\\
 \downarrow\uparrow \to \uparrow\uparrow:           \;\;\;\;                                 V_{31}&=&0    \nonumber\\
 \downarrow\uparrow \to \frac{1}{\sqrt{2}}(\uparrow\downarrow+\downarrow\uparrow): \;\;\;\;  V_{32}&=&\frac{1}{\sqrt{2}}q_z    \nonumber\\
\downarrow\uparrow \to \downarrow\downarrow:      \;\;\;\;                                   V_{33}&=&q_+   \nonumber\\
\downarrow\downarrow \to \uparrow\uparrow:        \;\;\;\;                                   V_{41}&=&0    \nonumber\\
\downarrow\downarrow \to \frac{1}{\sqrt{2}}(\uparrow\downarrow+\downarrow\uparrow): \;\;\;\; V_{42}&=& \frac{1}{\sqrt{2}}q_- \nonumber\\ 
\downarrow\downarrow \to \downarrow\downarrow:                            \;\;\;\;           V_{43}&=&-q_z   \nonumber\\   
\end{eqnarray}
\\
 Accounting for {\bf mechanism (b)} in  Eq.~(\ref{10.1})), $W_{ij}$:
\begin{eqnarray}
 \uparrow\uparrow \to \uparrow\uparrow:           \;\;\;\;                                  W_{11}&=&q_z    \nonumber\\
 \uparrow\uparrow \to \frac{1}{\sqrt{2}}(\uparrow\downarrow+\downarrow\uparrow): \;\;\;\;   W_{12}&=&\frac{1}{\sqrt{2}}q_+   \nonumber\\
 \uparrow\uparrow \to \downarrow\downarrow:                           \;\;\;\;              W_{13}&=&0    \nonumber\\  
 \uparrow\downarrow \to \uparrow\uparrow:                             \;\;\;\;              W_{21}&=&0    \nonumber\\
 \uparrow\downarrow \to \frac{1}{\sqrt{2}}(\uparrow\downarrow+\downarrow\uparrow): \;\;\;\; W_{22}&=&\frac{1}{\sqrt{2}}q_z    \nonumber\\
 \uparrow\downarrow \to \downarrow\downarrow:   \;\;\;\;                                    W_{23}&=&q_+    \nonumber\\
 \downarrow\uparrow \to \uparrow\uparrow:       \;\;\;\;                                    W_{31}&=&q_-    \nonumber\\
 \downarrow\uparrow \to \frac{1}{\sqrt{2}}(\uparrow\downarrow+\downarrow\uparrow): \;\;\;\; W_{32}&=&-\frac{1}{\sqrt{2}}q_z    \nonumber\\
\downarrow\uparrow \to \downarrow\downarrow:       \;\;\;\;                                 W_{33}&=&0    \nonumber\\
\downarrow\downarrow \to \uparrow\uparrow:      \;\;\;\;                                    W_{41}&=&0   \nonumber\\
\downarrow\downarrow \to \frac{1}{\sqrt{2}}(\uparrow\downarrow+\downarrow\uparrow):\;\;\;\; W_{42}&=&\frac{1}{\sqrt{2}}q_-   \nonumber\\ 
 \downarrow\downarrow \to \downarrow\downarrow:         \;\;\;\;                            W_{43}&=&-q_z    \nonumber\\   
\end{eqnarray}
\\

\section{$F'$ and $G'$ functions.}

The $F'$ and $G'$ functions in Eq.~(\ref{12.2}) are defined as:            

\begin{widetext}
\begin{align}
  &F'(P^0,P'^0,\vec{q},\omega_K,\vec{P},\vec{k})=\frac{1}{2\omega(\vec{q})}\frac{M_N}{E_N(\vec{P}-\vec{q}-\vec{k})}\frac{M_N}{E_N(-\vec{P}+\vec{q})}\frac{1}{\sqrt{s}-k^0-E_N(-\vec{P}+\vec{q})-E_N(\vec{P}-\vec{q}-\vec{k})+i\eps}\nonumber\\&\times\left\{\sum_{i=1,2}\frac{M_{\Lambda^*}^{(i)}}{E_{\Lambda^*}^{(i)}(\vec{P}-\vec{q})}\frac{(g^{(i)}_{\Lambda^*,K^-p})^2}{P^0-\omega_K(\vec{q})-E^{(i)}_{\Lambda^*}(\vec{P}-\vec{q})+i\frac{\Gamma_{\Lambda^*}^{(i)}}{2}}\frac{1}{P^0-\omega_K(\vec{q})-k^0-E_N(\vec{P}-\vec{q}-\vec{k})+i\eps}\right.\nonumber\\&+\left.\left(\frac{1}{P^0-E_{\Lambda^*}(\vec{P}-\vec{q})-\omega_K(\vec{q})+i\frac{\Gamma_{\Lambda^*}}{2}}+\frac{1}{P'^0-E_N(-\vec{P}+\vec{q})-\omega_K(\vec{q})+i\eps}\right)t_{K^-p,K^-p}(M_\mathrm{inv})\right\}\nonumber\\&G'(P^0,P'^0,\vec{q},\omega_K,\vec{P},\vec{k})=\frac{1}{2\omega(\vec{q})}\frac{M_N}{E_N(-\vec{P}-\vec{q}-\vec{k})}\frac{M_N}{E_N(\vec{P}+\vec{q})}\frac{1}{\sqrt{s}-k^0-E_N(\vec{P}+\vec{q})-E_N(-\vec{P}-\vec{q}-\vec{k})+i\eps}\nonumber\\&\times\left\{\sum_{i=1,2}\frac{M_{\Lambda^*}^{(i)}}{E_{\Lambda^*}^{(i)}(-\vec{P}-\vec{q})}\frac{g^{(i)}_{\Lambda^*,K^-p}g^{(i)}_{\Lambda^*,\pi^+\Sigma^-}}{P^0-\omega_\pi(\vec{q})-E^{(i)}_{\Lambda^*}(-\vec{P}-\vec{q})+i\frac{\Gamma_{\Lambda^*}^{(i)}}{2}}\frac{1}{P'^0-\omega_\pi(\vec{q})-k^0-E_N(-\vec{P}-\vec{q}-\vec{k})+i\eps}\right.\nonumber\\&\left.+\left(\frac{1}{P'^0-E_{\Lambda^*}(-\vec{P}-\vec{q})-\omega_\pi(\vec{q})+i\frac{\Gamma_{\Lambda^*}}{2}}+\frac{1}{P^0-E_N(\vec{P}+\vec{q})-\omega_\pi(\vec{q})+i\eps}\right)t_{K^-p,\pi^+\Sigma^-}(M'_\mathrm{inv}) \ \right\}\nonumber\\\label{eq:fs}\end{align}
\end{widetext}

In order to factorize the $t_{l,m}$ amplitudes, the mass and width of the heavier resonance at $1426$ MeV are taken in the terms $(P^0-E_{\Lambda^*}(\vec{P}-\vec{q})-\omega_K(q)+i\frac{\Gamma_{\Lambda^*}}{2})^{-1}$ of $F'$ and $(P'^0-E_{\Lambda^*}(-\vec{P}-\vec{q})-\omega_\pi(q)+i\frac{\Gamma_{\Lambda^*}}{2})^{-1}$ of $G'$. This can be done because these terms are very small compared to the other terms in the same bracket and this resonance is the one giving the largest contribution. 

\bibliography{biblioka}

\end{document}